# Surface structures of the magnetostrictive D0$_3$-Fe$_3$Ga(001)

Ricardo Ruvalcaba[1], J.P. Corbett[2*], Andrada-Oana Mandru[3], N. Takeuchi[1], Arthur R. Smith[4], J. Guerrero-Sanchez[1*]

[1]*Centro de Nanociencias y Nanotecnología, Universidad Nacional Autónoma de México, Ensenada, BC 22860, Mex.*
[2]*Department of Physics, the Ohio State University, Columbus, OH 43210, USA*
[3]*Empa, Swiss Federal Laboratories for Materials Science and Technology, Dübendorf CH-8600, Switzerland*
[4]*Nanoscale and Quantum Phenomena Institute—Department of Physics and Astronomy, Ohio University, Athens, OH 45701, United States of America*
[*]corbett.123@osu.edu, [*]guerrero@cnyn.unam.mx

## Abstract

First-principles total energy calculations and experimental measurements were performed to study the surface reconstructions of the magnetostrictive Fe$_3$Ga alloy. The magnetostrictive behavior was evaluated in the bulk by compressing and stretching its lattice parameter. Results demonstrate two thermodynamically stable surfaces, the 1×1 and 3×1 reconstructions. The 1×1 is an ideally FeGa terminated surface whereas the 3×1 is also FeGa terminated but it has a first-layer Fe atom substituted by a Ga atom every three unit-cells, forming stripe-like domain patterns. Tersoff–Hamann scanning tunneling microscopy simulations were obtained and compared with experimental results. We found good agreement between theory and experiment, in which the distance between rows is ~12.3 Å. The substrate-induced strain increases the stability of the 3×1 reconstruction. Here we have demonstrated that Ga/Fe atomic exchanges lead to the stripe-like domain patterns. Clarification of the atomic reconstructions present on the magnetostrictive Fe$_3$Ga alloys is an important step towards the understanding of its surfaces and poses this system as a potential candidate to be used as part of perpendicular magnetic tunnel junctions due to the existing perpendicular magnetic anisotropy effect when grown on different substrates.

# 1. Introduction

The study of magnetism and magnetic materials is not something new to the scientific community. Although magnetism can be considered as a well-studied phenomenon, there is current research on the topic regarding magnetic fluids, magneto electrochemistry, and magnetic materials' properties [1].

Two properties in particular - magnetostriction and its reciprocal effect magnetoelasticity - have a particularly wide variety of applications. The ability to change its magnetization when applied a strain and vice versa allows a particular material to be used to create wireless, passive sensors that can detect applied force [2], pressure [3], stress [4] and even strain [5]; all of which can then be applied in technological devices. Particularly in the medical field, magnetoelastic sensors have been used to measure the load on biomedical implants and devices [6], as well as to monitor the strain on a bone plate [7], the degradation of artificial bones in-vitro [8,9] and the tensile force on sutured wound sites [10], to mention some examples. But that's not it, by using electrodeposition as a means of producing coatings on magnetostrictive materials one can also create sensors for many different chemicals. Some of the chemicals that these sensors can successfully detect are carbon dioxide [11], ammonia [12], glucose [13], ricin [14], avidin [15] and staphylococcal enterotoxin B [16]. Additionally, because of its mechanical response to a magnetic signal, magnetostrictive materials may also be used as actuators for precision machining tools [17, 18, 19]. A changing magnetic field can be used in conjunction with a magnetostrictive material to produce vibrations. Such vibrators have already been looked upon [20, 21, 22] and they can be used in a varying range of applications. Anything from machining hard materials [23] or structural vibration control [24, 25] to making a motor [26, 27].

Among the magnetic alloys, galfenol ($Fe_{100-x}Ga_x$) and alfenol ($Fe_{100-x}Al_x$) are newer alloys that have more desirable mechanical properties than Terfenol-D ($Tb_xDy_{1-x}Fe_2$), i.e. the most commonly used magnetostrictive material in engineering [28]. Both galfenol and alfenol's most common structures have a BCC lattice, with <100> as the magnetic easy axis for magnetostriction and demonstrate sufficient ductility for the former sensor and actuator applications [29]. In the past couple of years, there have been many studies regarding galfenol's magnetic and mechanical properties [30, 31, 32, 33]. In one of them in particular, Andrada-Oana, et. al. [34] analyzed the FeGa growth to achieve a deeper understanding of the change in magnetic properties when depositing it on various substrates. Stripe-like domains as well as perpendicular magnetic anisotropy appeared when deposited on different substrates. Although interesting results emerged, no explanation of the atomic arrangements at the interface and surface has been assessed yet. Establishing the atomic arrangements of the possible surface reconstructions is a starting point to understand the interface formation and the change in magnetic properties due to the strain induced by the substrate since from a realistic point of view, the surface must be considered when analyzing the interface formation. The existence of atomic reconstructions at the surface may govern the way atomic arrangements at the interfaces are formed. In this paper, we aim to first describe the surface and to do that we have used density functional theory together with a thermodynamic stability analysis to describe the surface reconstruction responsible for the stripe-like patterns seen in the experimental STM images.

Results indicate two stable surface reconstructions: an ideal 1×1 FeGa terminated surface and a 3×1 FeGa terminated reconstruction with an atomic Ga/Fe exchange (a Ga atom takes an

Fe atom's place on the surface) every three unit-cells, leading to the formation of Ga channels, as evidenced in the calculated STM images, which perfectly agree with the experimental STM images. The magnetic properties of both surfaces are discussed as well. The rest of the paper is organized as follows: In section 2, we describe the methodology, section 3 is devoted to present and discuss the results, and finally, in section 4, we make some important conclusions and perspectives.

## 2. Methodology

Spin-polarized calculations were performed using the Plane-Wave Self-Consistent Field code as implemented in the Quantum ESPRESSO package [35]. The exchange-correlation energy was calculated based on the Generalized Gradient Approximation (GGA) by employing the Perdew–Burke–Ernzerhof (PBE) parametrization [36]. Ultrasoft scalar-relativistic pseudopotentials [37] were used on all of the calculations. By performing a minimum-energy optimization, the kinetic energy cutoff for wavefunctions and charge density were set to 45 and 360 Ry, respectively. The atomic relaxations and self-consistent calculations were performed using Brillouin-zone Monkhorst-Pack Γ-centered grids [38] of $10 \times 10 \times 10$ k-points for the bulk, and $10 \times 10 \times 1$ and $3 \times 10 \times 1$ k-points for the $(1 \times 1)$ and $(3 \times 1)$ surfaces, respectively. Methfessel-Paxton broadening [39] with a smearing width of $1 \times 10^{-2}$ Ry was used. The bulk structure was fully optimized and the lattice parameters used for the rest of the paper are $a, c = 4.10, 5.81$ Å for the $D0_3$ phase; which were in excellent agreement with the reported values $a, c = 4.10, 5.79$ Å. [40]. The improved tetrahedron method as implemented by Blöchl, et. al. [41] for

Brillouin zone integrations was used for total and projected Density Of States (DOS) calculations. The magnetic moment of each atom was analyzed for the $D0_3$ phase at strains ranging from -0.01 to 0.03 both iso and anisotropically. Surface Formation Energy (SFE) formalism adapted from Qian, Martin, and Chadi [42] was used to determine the most stable surface reconstruction. Finally, STM images as implemented by Tersoff-Hamann [43] were obtained for the most stable surfaces at 4 Å with a bias voltage of +2.0 V, as utilized in [34].

## 3. Results and Discussions

### 3.1. Structural, electronic and magnetic properties of the $Fe_3Ga$ bulk structure

In [34] it was hypothesized that the observed $Fe_3Ga$ film deposited onto MnN was a mixture of ordered phases close to the Ga concentration used (%at. Ga ~ 23.6). By referring to the $Fe_3Ga$ phase-diagram [44], these phases could either be $B_2$, $D0_3$, or a mixture of ordered $L1_2$ with disordered $A_2$ phases. The $B_2$-phase was discarded due to its instability below 650°C [40] since the surfaces were obtained by Molecular Beam Epitaxy (MBE) at temperatures between 360 and 420 °C. Formation of equilibrium FCC ordered $L1_2$ phase starts after four hours annealing at 350 °C [45]. The film thickness reported on [34] was approximately 4.8 ± 0.4 Å and the typical MBE growth rate is approximately 1 Å/s (one atomic layer per second) [46]. Therefore, we may estimate that growth time was much less than the required to form the $L1_2$ phase, leaving the $D0_3$-phase as the main candidate to explain the observed row structure.

The $Fe_3Ga$ bulk structure was studied for the $D0_3$ phase, which has a face-centered tetragonal (FCT) structure [40], see Figure 1a. Non-magnetic (NM), antiferromagnetic (AFM), and ferromagnetic (AFM) configurations were tested during the structural relaxation. The bulk energies were compared and the FM order was found to be the most stable one, whereas the NM and AFM are less stable by 1.22 eV/Fe$_3$Ga unit and 0.25 eV/Fe$_3$Ga unit, respectively (Figure 1b). These results agree with the literature, as it is well known that Fe-Ga alloys behave as ferromagnets [47].

As for the electronic properties, the calculated spin-polarized DOS in Fig. 1c shows that most of the spin-up states are located below the Fermi level, in clear asymmetry with the spin-down states, therefore confirming the ferromagnetic behavior. The projected DOS show that the main contribution to the states near the Fermi level comes from the Fe d-orbitals (particularly the d$_{xz,yz}$ orbitals), with almost zero-contribution from Ga atoms to such states. Note that the D0$_3$ structure is formed by alternating Fe and FeGa layers, which leads to the presence of two kinds of Fe atoms, Fe(1) and Fe(2). Fe(1) is related to the FeGa layer and Fe(2) to the Fe layer. Note the hybridization between Fe(1)-d and Ga-p orbitals (green dashed and red lines, respectively). These results also come in agreement with the literature [40]. Since Fe atoms have different chemical surroundings, a difference in the magnetic moments is to be expected.

We now proceed to analyze the magnetostrictive behavior of the D0$_3$ (Fig. 1d) considering that in [34] Fe$_3$Ga is grown on MgO, MnN, and Mn$_3$N$_2$ substrates. Such substrates will generate different lattice mismatches (between 2.5% and 3%). The D0$_3$ was strained up until the 3% strain value, which includes all the possible substrate-induced strains. For the sake of completeness we compressed the system in 1% to study the magnetic moments' behavior. Isotropic and anisotropic (in-plane) strains were analyzed. It is clear in both that the magnetic moments increase linearly with the strain within the examined region. Notoriously, Fe(1)'s magnetic moment is always greater than Fe(2)'s for the examined region. This phenomenon is also associated with the weak Fe-Ga hybridization on the Fe(2). The magnetostrictive behavior of $Fe_3Ga$ alloys could be due to the presence of nonbinding Fe d-states around the Fermi level (as shown in Fig. 1c). To see a more detailed explanation of this phenomenon please see Zhang and Wu [48]. The different row-like patterns experimentally found on a sample deposited onto AFM MnN (which induces a strain of ~3%) [34] are not formed due to the magnetostriction of the $Fe_3Ga$ alloy nor to the lattice mismatches induced by the different substrates. It is more likely to be a series of surface reconstructions. The assessment of this idea is discussed in the following sections.

## 3.2. Stability of the $Fe_3Ga(001)$ surfaces through the SFE formalism

In order to consider all the possible surface terminations and surface reconstructions, we used Fe-only terminated 2×1 and FeGa-terminated 3×1 surface unit cells. As previously mentioned, the surfaces were grown experimentally in [34] via MBE at temperatures between 360 and 420 °C. Taking into consideration all the possible defects that may arise, both surface reconstructions had either substituted or adsorbed atoms, as illustrated in Fig. 2. A and B are the ideal non-reconstructed surfaces with 2×1 and 3×1 periodicities, respectively. Labels $a_x$ and $s_x$ (x=1,2,3) stand for adsorption and substitution positions on the surfaces, respectively. The nomenclature for the surface reconstructions goes as follows: (reconstruction periodicity).(metal added).(position of adsorption/substitution). Hence, reconstructions $A.Fe.a_1$ and $A.Fe.a_2$ have 2×1 periodicities with an Fe atom adsorbed in positions $a_1$ and $a_2$, respectively. The same goes for $A.Ga.a_1$ and $A.Ga.a_2$ but with Ga atoms. Reconstruction $A.Ga.s_1$ has a 2×1 periodicity with a Ga atom substituting the Fe atom in position $s_1$. Similarly, reconstructions $B.Fe.a_3$ and $B.Ga.a_3$ have 3×1 periodicities with Fe/Ga atoms adsorbed in position $a_3$, respectively. Finally, reconstructions B.Fe.$s_2$ and B.Ga.$s_3$ have a 3×1 periodicities with Fe/Ga atoms substituted in positions $s_2$ and $s_3$, respectively. Notice that adsorption positions can be filled with both metals whilst substitutions only with one. This happens because one atom replacing another identical one leaves the surface intact. Inversion symmetry was kept on every cell, therefore having two equivalent surfaces. Empty spaces of ~11 Å separate the surfaces in order to avoid interaction with their equivalent surface.

Since we are treating with a different number of atoms in each surface reconstruction, total energies of these systems are not comparable, so the stability of them needs to be assessed by using first-principles thermodynamics [42, 49]. To apply this formalism we must consider thermal equilibrium between vacuum (reservoir), surface, and bulk. So each change in energy due to the addition or removal of atoms is registered by their corresponding chemical potentials. For condensed phases, chemical potentials are defined as the total energy per atom of their most stable bulk structure, the one existing in nature [42].

This theory was applied to compare the different surfaces because the stoichiometry was different among them. The expression for the SFE in the proposed system is found to be:

$$SFE = \frac{1}{2A}\left[E_{slab} - \mu_{bulk}\left(\frac{1}{6}n_{Fe} + \frac{1}{2}n_{Ga}\right) - \Delta\mu\left(\frac{1}{6}n_{Fe} - \frac{1}{2}n_{Ga}\right)\right] \quad (1)$$

where $A$ is the surface area, $E_{slab}$ is the final energy of the surface, $n_{Fe}$ and $n_{Ga}$ are the numbers of atoms of each element present in the surface, and $\mu_{Fe}$, $\mu_{Ga}$, and $\mu_{bulk}$ are the chemical potentials of the Fe, Ga and D0$_3$-Fe$_3$Ga bulk structures, respectively. The SFE was plotted from Ga-rich ($-\Delta H_f \leq \Delta\mu \leq 0$) to Fe-rich conditions ($0 \leq \Delta\mu \leq \Delta H_f$), where $\Delta H_f$, which defines the allowed growth limits, is the Fe$_3$Ga formation enthalpy, defined as $\Delta H_f = 3\mu_{Fe}+\mu_{Ga}-\mu_{bulk}$. The formation enthalpy was found to be 0.188 eV/Fe$_3$Ga unit (D0$_3$ phase), which has an excellent agreement with the reported values of 0.376 ± 0.58 [50] and 0.416 ± 0.328 eV/Fe$_3$Ga unit [51]. However, reported values are not as precise as one would hope, making the equality of signs between calculated and reported values the most significant parameter. A positive value under this definition means that the D0$_3$ phase is more stable than its constituents (Fe and Ga chemical potentials) and otherwise if negative.

Results of the stability analysis are plotted in Fig 3a, the lowest energy values are related to the most stable surfaces. Note that the Fe terminated surface (A surface) is not stable under any range of chemical potential. On the other hand, the stable surfaces are the FeGa ideally terminated surface (B surface) in a 1×1 simple reconstruction (for Fe rich conditions) and the 3×1 reconstruction induced in the B surface by making it rich in Ga. To do this, an Fe atom of the first layer has been exchanged by a Ga atom, which leads to a stable structure, the B.Ga.s$_3$ reconstruction. This surface is stable from Ga intermediate to Ga rich conditions, see Fig 3a. The stability range of this reconstruction is expected since we are generating surfaces with a certain lack of Fe on the first layer. No 2×1 reconstructions were stable. Note that Fe adsorbed on the Fe terminated surface generates the less stable reconstructions, Fe adsorbed on the FeGa terminated surface diminishes its stability. Fe substitution on the FeGa-terminated surface is also unstable. Note that the structures stabilize as they increase their Ga content. So, it is to be expected that Ga layers could be engineered in this system. Similar results reporting stable Ga layers have recently appeared in the literature [52].

In Fig 3b it is plotted the effect of the maximum induced strain due to the substrate. Generally speaking, in an experimental growth procedure the film experiences an elongation or contraction of its lattice parameter in order to match the one imposed by the substrate. Here we simulate the maximum strain induced by the MnN substrate on the $D0_3$-$Fe_3Ga$ and analyze how it affected the stability of the favored models. Note that the lattice mismatch between MnN and $D0_3$-$Fe_3Ga$ is of ~3%, small enough to assure epitaxial growth with the minimum quantity of defects at the interface. The interface analysis considering surface effects is a matter of a complete study and is beyond the scope of this work. Induced strain reduces the growth limit imposed by the formation enthalpy, from 0.188 to 0.156 eV/$Fe_3Ga$ unit when under ~3% of strain. Stable models were also fully optimized with the new lattice parameter (~3% larger). Note that the 1×1 FeGa terminated surface experiences a decrease in stability, being achievable just for very rich Fe conditions. On the other hand, the 3×1 $B.Ga.s_3$ structure gains stability and is favored for almost all growth conditions. Our results indicate that the extra Ga atoms on the 3×1 surface help to stabilize the 3×1 reconstruction under such strain and it may be the reason why FeGa clean surface only appears as small patches all over the STM images.

*3.3 Scanning tunneling microscopy analysis: DFT and experimental data*

Since our main goal in this manuscript is to explain the row-structure appearing in the $D0_3$-$Fe_3Ga$ thin films, we proceed to compare the theoretical Tersoff-Hamann STM images with the experimental results. In the Fe-Ga phase diagram, films containing 15 % to 23 % Ga grow in a mixed phase of disordered $A_2$ and ordered tetragonal $D0_3$, above 23% phases of $B_2$ and $D0_3$ coexist [34,44]. The samples were prepared under Ga concentrations (~23%) which would be predominately the D0$_3$ phase as explained before. From the bulk crystalline phase diagram, accessing the different phases depends critically on the formation temperatures and quenching rates [44,53]. While these formation windows tend to be similar for thin-film growths, it is worthwhile to note in the theoretical phase diagram the D0$_3$ is the ground state crystallization for this concentration [53].

As mentioned before, it was found from STM observations a row-structure formation on Fe-Ga thin films with a ~23% Ga concentration [34]. A well-ordered (3×1) structure with a row spacing of 12.8 ± 0.5 Å (short rows) and a dislocated (4.5×1) structure with a row spacing of 18.6 ± 0.5 Å (wide rows) were observed, see Figure 4(a). Although conclusions regarding the surface structures were not drawn. Through the use of DFT and TH-STM images we are able to untangle the makeup of the observed row-structure.

Since the A$_2$ structure is α-Fe with random Ga substitutions, it would seem unlikely that an ordered surface structure would emerge. If this row-structure does arise from an A$_2$ phase it would be a (6×1) and (9×1) surface structure for the short and wide rows respectively. This would be a scenario with large surface unit cells and many equivalent unoccupied substitution sites between the rows, leading one to suspect this as an unlikely scenario given the high concentration of Ga and the well-ordered surface.

In turn, this leads us to investigate the D0$_3$ phase as the candidate to host the row-structure. There are two possible ideal terminations, an FeGa termination (B surface) and an Fe termination (A surface) as shown in Figure 2. It is found from the SFE calculations in Figure 3a, the ideal Fe surface has higher energy than the ideal FeGa surface (labeled A and B respectively in the SFE plot). There is a clear trend in SFE data, if you start with ideal Fe (or

FeGa) termination and add more Fe it becomes even more unfavorable, however by adding Ga the favorability improves.

With the SFE trend in mind, starting with the Fe layer and adding an Fe to (6×1) or (9×1) would be unfavorable as adding more Fe increases energy. In the case of a Ga replacement, this would improve the energy favorability, but so little Ga is being added that it would be approximately equal to the ideal Fe SFE. Furthermore, for more densely Ga substituted Fe surfaces the SFE is still significantly higher than the ideal and substituted FeGa terminations. Therefore, we focus on the FeGa termination and substitutions that could give rise to the observed row-structure.

Both Fe and Ga substitutions of the ideal FeGa termination are considered when modeling the well-ordered (3×1) structure. A Ga substituted (3×1) structure is found with lower energy than the ideal (1×1) FeGa terminated surface, see Figure 1(c-d). Theoretical TH-STM images of the two lowest energy (3×1) and (1×1) structures are computed where the (3×1) produces a pronounced row-structure, see Figure 1(d). Fast Fourier Transforms (FFT) of the experimental and theoretical STM images are shown in Figure 1(e-g). A line of dense k-points corresponds to the (3×1) row-structure, as indicated by the green arrows in figure 1(e,f). While FFT of the (3×1) TH-STM image has additional weaker k-points in a square array (see blue arrow) corresponding to the underlying FeGa surface structure. The FFT of the experimental STM image lacks the additional weaker k-points, however, this is to be expected from the real-space image as only possible weak hints of atomic resolution are present. The FFT of the ideally terminated (1×1) FeGa surface shows a square array of 4 k-points indicated by the blue arrow in Figure 1(g). Overall, good agreement is achieved between the experimental STM and

TH-STM images enabling us to determine the (3×1) row-structure as a Ga-substituted FeGa terminated surface.

Further work experimentally and theoretically is needed to untangle the details of the (4.5×1) row-structure. We suspect this is a Ga-substitution given the SFE trend and that wide rows can coexist within short rows as seen in Figure 1(a) marked by S W S. There are additionally unresolved regions, as indicated by the dotted light blue boarders, where the (3×1) structure did not completely substitute the termination in Ga, this would be the ideal terminated FeGa surface although we lack the resolution to resolve the square lattice. Further work should be done to increase the resolution of the row-structures and unresolved regions to pin down additional details that cannot be achieved from this analysis.

## 4. Conclusions

This paper presented *ab initio* calculations together with experimental STM measurements to study the $Fe_3Ga(001)$ surfaces. The magnetostrictive behavior of the bulk $D0_3$ phase was studied when expanding the unit cell isotropically and anisotropically (in-plane), and it was associated with a weak Fe-Ga hybridization on the Fe layers. An FeGa terminated 1×1 surface and a row-like 3×1 reconstruction -FeGa terminated surface with a first-layer Fe exchanged by Ga each three unit cells- are the thermodynamically stable structures. Theoretical and experimental STM images were compared and found a close agreement, showing a distance between the stripe-like domains of ~12.3 Å. When strain is induced to the stable surfaces, the 3×1 reconstruction turns to be the only favored demonstrating that extra Ga atoms help to stabilize the stripe-like domain patterns. Our study is a step towards the understanding of the magnetostrictive $Fe_3Ga$ atomic reconstructions, posing this system as a candidate to be used as part of magnetic heterojunctions with potential applications in spintronics devices.

## 5. Acknowledgments

The authors would like to thank DGAPA-UNAM projects IN100219 and IA100920, and CONACYT grant A1-S-9070 of the Call of Proposals for Basic Scientific Research 2017-2018 for partial financial support. We acknowledge the DGTIC-UNAM supercomputing center project LANCAD-UNAM-DGTIC-368, and *Laboratorio Nacional de Supercómputo del Sureste de México*, CONACYT - member of the network of national laboratories - and the technical support provided by E. Murillo and Aldo Rodriguez-Guerrero.

## 6. References

[1] J. Coey, Magnetism and magnetic materials. Cambridge: Cambridge University Press, 2010, pp. 542-565.
[2] B. Pereles, A. DeRouin and K. Ghee Ong, "A Wireless, Passive Magnetoelastic Force–Mapping System for Biomedical Applications", Journal of Biomechanical Engineering, vol. 136, no. 1, 2013.
[3] E. Tan, B. Pereles and K. Ong, "A Wireless Embedded Sensor Based on Magnetic Higher Order Harmonic Fields: Application to Liquid Pressure Monitoring", IEEE Sensors Journal, vol. 10, no. 6, pp. 1085-1090, 2010.
[4] D. Kouzoudis and C. Grimes, "The frequency response of magnetoelastic sensors to stress and atmospheric pressure", Smart Materials and Structures, vol. 9, no. 6, pp. 885-889, 2000.
[5] N. Oess, B. Weisse and B. Nelson, "Magnetoelastic Strain Sensor for Optimized Assessment of Bone Fracture Fixation", IEEE Sensors Journal, vol. 9, no. 8, pp. 961-968, 2009.
[6] B. Pereles, A. DeRouin and K. Ghee Ong, "A Wireless, Passive Magnetoelastic Force–Mapping System for Biomedical Applications", Journal of Biomechanical Engineering, vol. 136, no. 1, 2013.
[7] Y. Tan, J. Hu, L. Ren, J. Zhu, J. Yang and D. Liu, "A Passive and Wireless Sensor for Bone Plate Strain Monitoring", Sensors, vol. 17, no. 11, p. 2635, 2017.

[8] K. Yu, L. Ren, Y. Tan and J. Wang, "Wireless Magnetoelasticity-Based Sensor for Monitoring the Degradation Behavior of Polylactic Acid Artificial Bone In Vitro", Applied Sciences, vol. 9, no. 4, p. 739, 2019.

[9] L. Ren, K. Yu and Y. Tan, "Monitoring and Assessing the Degradation Rate of Magnesium-Based Artificial Bone In Vitro Using a Wireless Magnetoelastic Sensor", Sensors, vol. 18, no. 9, p. 3066, 2018.

[10] A. DeRouin, N. Pacella, C. Zhao, K. An and K. Ong, "A Wireless Sensor for Real-Time Monitoring of Tensile Force on Sutured Wound Sites", IEEE Transactions on Biomedical Engineering, vol. 63, no. 8, pp. 1665-1671, 2016.

[11] Q. Cai, A. Cammers-Goodwin and C. Grimes, "A wireless, remote query magnetoelastic $CO_2$ sensor", Journal of Environmental Monitoring, vol. 2, no. 6, pp. 556-560, 2000.

[12] Q. Cai, M. Jain and C. Grimes, "A wireless, remote query ammonia sensor", Sensors and Actuators B: Chemical, vol. 77, no. 3, pp. 614-619, 2001.

[13] K. Ong et al., "Magnetism-Based Remote Query Glucose Sensors", Sensors, vol. 1, no. 5, pp. 138-147, 2001.

[14] C. Ruan et al., "A Magnetoelastic Ricin Immunosensor", Sensor Letters, vol. 2, no. 2, pp. 138-144, 2004.

[15] C. Ruan, K. Zeng, O. Varghese and C. Grimes, "A magnetoelastic bioaffinity-based sensor for avidin", Biosensors and Bioelectronics, vol. 19, no. 12, pp. 1695-1701, 2004.

[16] C. Ruan, K. Zeng, O. Varghese and C. Grimes, "A staphylococcal enterotoxin B magnetoelastic immunosensor", Biosensors and Bioelectronics, vol. 20, no. 3, pp. 585-591, 2004.

[17] D. Davino, C. Natale, S. Pirozzi and C. Visone, "Phenomenological dynamic model of a magnetostrictive actuator", Physica B: Condensed Matter, vol. 343, no. 1-4, pp. 112-116, 2004.

[18] C. Hong, "Application of a magnetostrictive actuator", Materials & Design, vol. 46, pp. 617-621, 2013.

[19] S. Karunanidhi and M. Singaperumal, "Design, analysis and simulation of magnetostrictive actuator and its application to high dynamic servo valve", Sensors and Actuators A: Physical, vol. 157, no. 2, pp. 185-197, 2010.

[20] L. Emory, "Magnetostrictive vibrator", US2249835A, 1958.

[21] P. Washington, "Magnetostrictive vibrator", US1882397A, 1949.

[22] T. Ueno, E. Summers, M. Wun-Fogle and T. Higuchi, "Micro-magnetostrictive vibrator using iron–gallium alloy", Sensors and Actuators A: Physical, vol. 148, no. 1, pp. 280-284, 2008.

[23] L. Balamuth, "Magnetostrictive vibrator for high frequency machining of hard materials", US3471724A, 1969.

[24] S. Moon, C. Lim, B. Kim and Y. Park, "Structural vibration control using linear magnetostrictive actuators", Journal of Sound and Vibration, vol. 302, no. 4-5, pp. 875-891, 2007.

[25] F. Braghin, S. Cinquemani and F. Resta, "A model of magnetostrictive actuators for active vibration control", Sensors and Actuators A: Physical, vol. 165, no. 2, pp. 342-350, 2011.
[26] F. Abbot, "Magnetostrictive vibration motor", US3470402A, 1969.
[27] F. Claeyssen, N. Lhermet, R. Le Letty and P. Bouchilloux, "Actuators, transducers and motors based on giant magnetostrictive materials", Journal of Alloys and Compounds, vol. 258, no. 1-2, pp. 61-73, 1997.
[28] A. Clark, "Chapter 7 Magnetostrictive rare earth-Fe2 compounds", Handbook of Ferromagnetic Materials, vol. 1, p. 531, 1980.
[29] J. Park, S. Na, G. Raghunath and A. Flatau, "Stress-anneal-induced magnetic anisotropy in highly textured Fe-Ga and Fe-Al magnetostrictive strips for bending-mode vibrational energy harvesters", AIP Advances, vol. 6, no. 5, p. 056221, 2016.
[30] E. Summers, T. Lograsso and M. Wun-Fogle, "Magnetostriction of binary and ternary Fe–Ga alloys", Journal of Materials Science, vol. 42, no. 23, pp. 9582-9594, 2007.
[31] A. Clark, M. Wun-Fogle, J. Restorff and T. Lograsso, "Magnetostrictive Properties of Galfenol Alloys Under Compressive Stress", MATERIALS TRANSACTIONS, vol. 43, no. 5, pp. 881-886, 2002.
[32] E. Summers, T. Lograsso, J. Snodgrass and J. Slaughter, "Magnetic and mechanical properties of polycrystalline Galfenol", Smart Structures and Materials 2004: Active Materials: Behavior and Mechanics, vol. 5387, 2004.
[33] G. Petculescu, K. Hathaway, T. Lograsso, M. Wun-Fogle and A. Clark, "Magnetic field dependence of galfenol elastic properties", Journal of Applied Physics, vol. 97, no. 10, 2005.
[34] A. Mandru et al., "Magnetostrictive iron gallium thin films grown onto antiferromagnetic manganese nitride: Structure and magnetism", Applied Physics Letters, vol. 109, no. 14, 2016.
[35] P. Giannozzi et al., "QUANTUM ESPRESSO: a modular and open-source software project for quantum simulations of materials", Journal of Physics: Condensed Matter, vol. 21, no. 39, 2009.
[36] J. Perdew, K. Burke and M. Ernzerhof, "Generalized Gradient Approximation Made Simple", Physical Review Letters, vol. 77, no. 18, pp. 3865-3868, 1996.
[37] A. Rappe, K. Rabe, E. Kaxiras and J. Joannopoulos, "Optimized pseudopotentials", Physical Review B, vol. 41, no. 2, pp. 1227-1230, 1990.
[38] H. Monkhorst and J. Pack, "Special points for Brillouin-zone integrations", Physical Review B, vol. 13, no. 12, pp. 5188-5192, 1976.
[39] M. Methfessel and A. Paxton, "High-precision sampling for Brillouin-zone integration in metals", Physical Review B, vol. 40, no. 6, pp. 3616-3621, 1989. Available: 10.1103/physrevb.40.3616 [Accessed 23 May 2020].
[40] R. Wu, "Origin of large magnetostriction in FeGa alloys", Journal of Applied Physics, vol. 91, no. 10, p. 7358, 2002.
[41] P. Blöchl, O. Jepsen and O. Andersen, "Improved tetrahedron method for Brillouin-zone integrations", Physical Review B, vol. 49, no. 23, pp. 16223-16233, 1994.

[42] G. Qian, R. M. Martin, and D. J. Chadi, "First-principles study of the atomic reconstructions and energies of Ga- and As-stabilized GaAs(100) surfaces", Phys. Rev. B, vol. 38, no. 11, pp. 7649, 1988.

[43] J. Tersoff and D. Hamann, "Theory of the scanning tunneling microscope", Physical Review B, vol. 31, no. 2, pp. 805-813, 1985.

[44] O. Goldbeck, IRON-Binary Phase Diagrams. Berlin, Heidelberg: Springer Berlin Heidelberg, 1982, p. 41.

[45] I. Golovin, V. Palacheva, A. Bazlov, J. Cifre and J. Pons, "Structure and anelasticity of Fe3Ga and Fe3(Ga,Al) type alloys", Journal of Alloys and Compounds, vol. 644, pp. 959-967, 2015.

[46] D. Leadly, "MBE - Molecular Beam Epitaxy", 2009. [Online]. Available: https://warwick.ac.uk/fac/sci/physics/current/postgraduate/regs/mpagswarwick/ex5/growth/pvd/. [Accessed: 15- May- 2020].

[47] C. J. Quinn, "The fabrication and analysis of the magnetic and crystallographic properties of Fe-rich (Fe x Ga 1-x) Galfenol alloys", Ph.D. thesis, Abbrev. Dept., University of Salford, 2012. Accessed on: 04/23/2020. [Online]. Available: http://usir.salford.ac.uk/id/eprint/28420/

[48] Y. Zhang and R. Wu, "Mechanism of Large Magnetostriction of Galfenol", IEEE Transactions on Magnetics, vol. 47, no. 10, pp. 4044-4049, 2011.

[49] A. Mandru, J. Pak, A. Smith, J. Guerrero-Sanchez and N. Takeuchi, "Interface formation for a ferromagnetic/antiferromagnetic bilayer system studied by scanning tunneling microscopy and first-principles theory", Physical Review B, vol. 91, no. 9, 2015.

[50] K. Persson, "Materials Data on GaFe3 (SG:225) by Materials Project", 2014. [Online]. Available: https://materialsproject.org/materials/mp-672661/ [Accessed: 15- May- 2020].

[51] J. E. Saal, S. Kirklin, M. Aykol, B. Meredig, and C. Wolverton, "Materials Data on GaFe3 (ID=18835) by OQMD", 2013. [Online]. Available: http://oqmd.org/materials/entry/18835 [Accessed: 15- May- 2020].

[52] M. Moreno-Armenta, J. Corbett, R. Ponce-Perez and J. Guerrero-Sanchez, "A DFT study on the austenitic Ni2MnGa (001) surfaces", Journal of Alloys and Compounds, vol. 836, p. 155447, 2020.

[53] M. Matyunina et al., "Phase diagram of magnetostrictive Fe-Ga alloys: insights from theory and experiment", Phase Transitions, vol. 92, no. 2, pp. 101-116, 2018. Available: 10.1080/01411594.2018.1556268 [Accessed 24 May 2020].

## 7. Figures

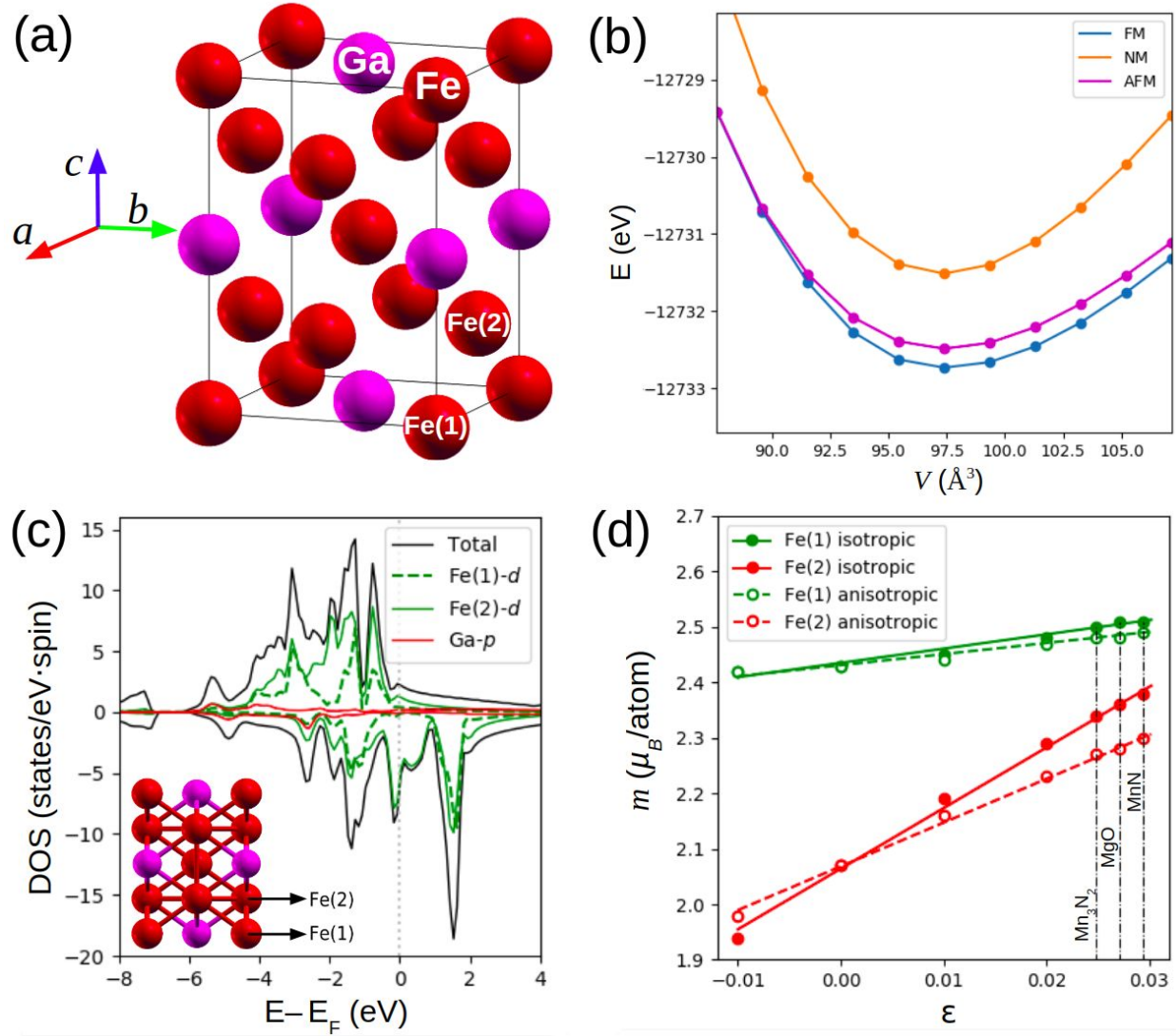

**Fig 1.** Results from the bulk structure calculations. (a) Atomic structure of the magnetostrictive $D0_3$-$Fe_3Ga$ alloy. Red (magenta) spheres represent the Fe (Ga) atoms. Crystallographic axes are added for the sake of spatial orientation. (b) Energy vs volume plot. Blue, magenta, and orange correspond to Ferromagnetic, Antiferromagnetic, and Non-magnetic structures, respectively. (c) Total and projected Density Of States of the most stable ferromagnetic structure. Black, dashed green, green, and red represent the total density of states, Fe(1)-d orbitals, Fe(2)-d orbitals, and Ga-p orbitals contributions to the DOS, respectively. (d) Fe magnetic moments vs strain plots. Continuous and dashed lines correspond to isotropic and in-plane applied strain, respectively. Perpendicular dot-dash-dot lines indicate the strains applied to $D0_3$-$Fe_3Ga$ by different substrates: 2.48%, 2.71%, and 2.94% for $Mn_3N_2$, MgO, and MnN, respectively.

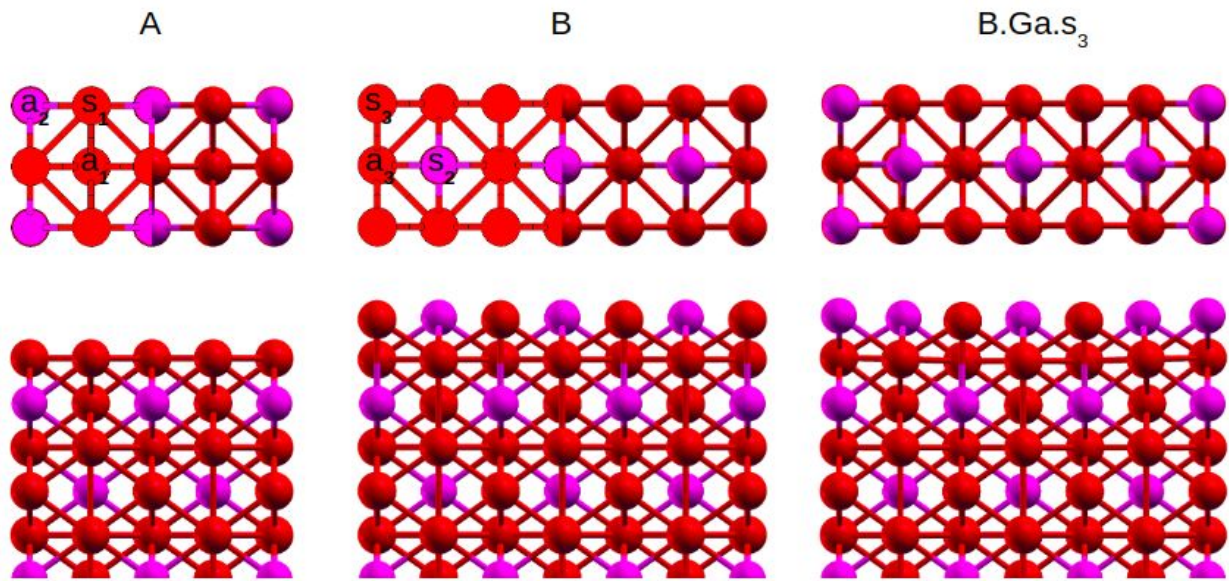

**Fig 2.** Top and front views of the proposed surface terminations and the most stable reconstruction. A is an ideal Fe-only terminated (1×1) surface. B is an ideal FeGa-terminated (1×1) reconstruction. Positions $a_1$, $a_2$, $a_3$ stand for adsorption sites and $s_1$, $s_2$, $s_3$ stand for atomic substitutions. Configuration B.Ga.$s_3$ is a B surface with a Ga substitution in position $s_3$.

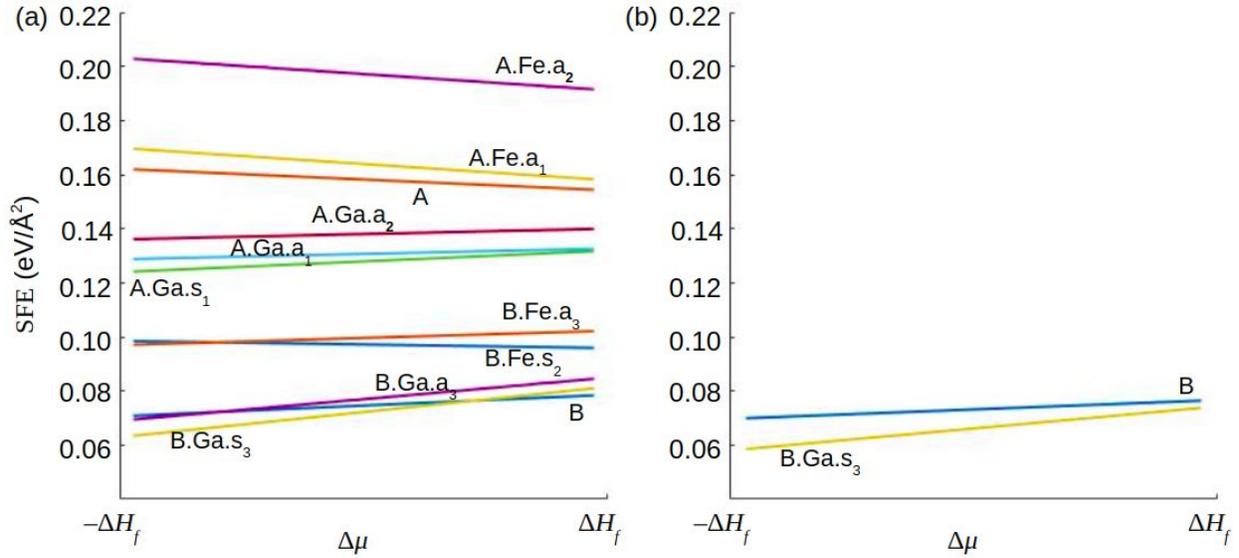

**Fig 3.** Surface Formation Energies as defined in Eq. (1) for (a) the proposed $Fe_3Ga(001)$ surfaces. A and B are the ideal non-reconstructed surfaces with 2×1 and 3×1 periodicities, respectively. Labels $a_x$ and $s_x$ (x=1,2,3) stand for adsorption and substitution positions on the surfaces, respectively. The nomenclature for the surface reconstructions goes as follows: (reconstruction periodicity).(metal added).(position of adsorption/substitution). Hence, reconstructions $A.Fe.a_1$ and $A.Fe.a_2$ have 2×1 periodicities with an Fe atom adsorbed in positions $a_1$ and $a_2$, respectively. The same goes for $A.Ga.a_1$ and $A.Ga.a_2$ but with Ga atoms. Reconstruction $A.Ga.s_1$ has a 2×1 periodicity with a Ga atom substituting the Fe atom in position $s_1$. Similarly, reconstructions $B.Fe.a_3$ and $B.Ga.a_3$ have 3×1 periodicities with Fe/Ga atoms adsorbed in position $a_3$, respectively. Finally, reconstructions $B.Fe.s_2$ and $B.Ga.s_3$ have a 3×1 periodicities with Fe/Ga atoms substituted in positions $s_2$ and $s_3$, respectively. (b) effect of the maximum strain on the stability as induced by a MnN substrate (~3%) for the most stable models only. Formation enthalpy experiences a reduction of 8 meV/atom. Lower yellow and blue lines correspond to $B.Ga.s_3$ and B models, respectively.

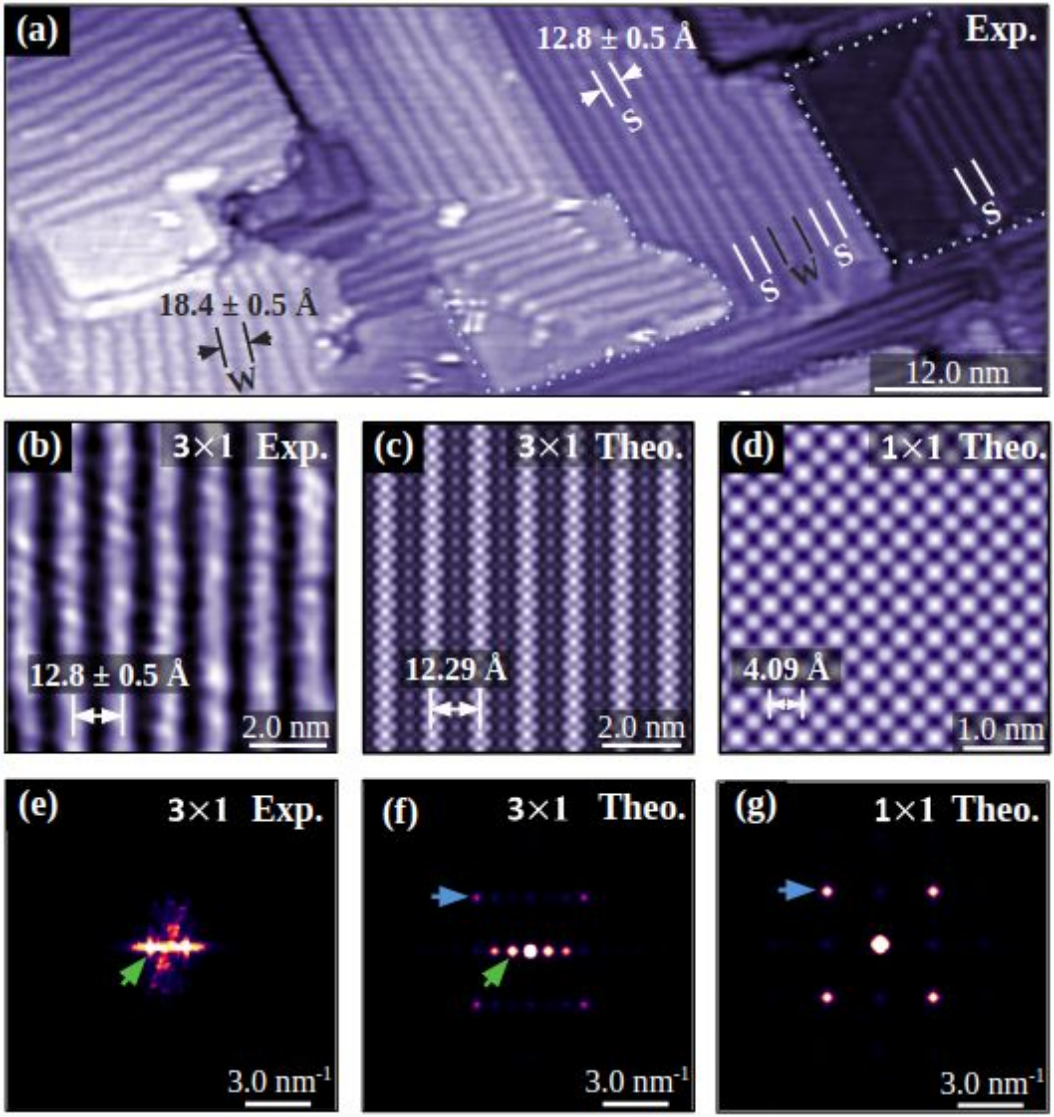

**Fig 4.** Experimental and theoretical STM images of the D0$_3$-Fe$_3$Ga surface. (a) Experimental STM image showing two row-structures, a well-ordered set of short rows (3×1) are indicated with a white S and a set of wide rows (4.5×1) are indicated by gray W. Unresolved regions are marked by a light blue dotted line. Atomic contrast was enhanced by Laplacian filtering. (b) Zoomed view of the experimental (3×1) row-structure. (c) Tersoff-Hamann-STM image of a (3×1) Ga substituted ideal FeGa termination layer of the Fe$_3$Ga crystal showing a row-structure in good agreement with the experiment. (c) Tersoff-Hamann-STM image of the ideal (1×1) FeGa termination. (e-g) FFT of the experimental and theoretical STM images in (b-d) respectively. A dense line of k-points indicates the row-structure (green arrows), while the (1×1) lattice can be seen as a square lattice of k-points (blue arrows).